# A Survey of Routing Attacks and Security Measures in Mobile Ad-Hoc Networks

Sudhir Agrawal, Sanjeev Jain, Sanjeev Sharma

**Abstract** – Mobile ad hoc networks (MANETs) are a set of mobile nodes which are self-configuring and connected by wireless links automatically as per the defined routing protocol. The absence of a central management agency or a fixed infrastructure is a key feature of MANETs. These nodes communicate with each other by interchange of packets, which for those nodes not in wireless range goes hop by hop. Due to lack of a defined central authority, securitizing the routing process becomes a challenging task thereby leaving MANETs vulnerable to attacks, which results in deterioration in the performance characteristics as well as raises a serious question mark about the reliability of such networks. In this paper we have attempted to present an overview of the routing protocols, the known routing attacks and the proposed countermeasures to these attacks in various works.

——————————— ◆ ———————————

## 1 INTRODUCTION

Mobile Ad hoc NETwork (MANET) [1] is a set of mobile devices (nodes), which over a shared wireless medium communicate with each other without the presence of a predefined infrastructure or a central authority. The member nodes are themselves responsible for the creation, operation and maintenance of the network. Each node in the MANET is equipped with a wireless transmitter and receiver, with the aid of which it communicates with the other nodes in its wireless vicinity. The nodes which are not in wireless vicinity, communicate with each other hop by hop following a set of rules (routing protocol) for the hopping sequence to be followed.

The chief characteristics and challenges of the MANETs [2] can be classified as follows:

*Cooperation:*
If the source node and destination node are out of range with each other then the communication between them takes place with the cooperation of other nodes such that a valid and optimum chain of mutually connected nodes is formed. This is known as multi hop communication. Hence each node is to act as a host as well as a router simultaneously.

*Dynamism of Topology:*
The nodes of MANET are randomly, frequently and unpredictably mobile within the network.[3] These nodes may leave or join the network at any point of time, thereby significantly affecting the status of trust among nodes and the complexity of routing. Such mobility entails that the topology of the network as well as the connectivity between the hosts is unpredictable. So the management of the network environment is a function of the participating nodes.

———————————————

- Sudhir Agrawal is with the Truba Institute of Engineering & Information Technology, Bhopal, India.
- Sanjeev Jain. is with Madhav Institute of Technology & Science, Gwalior, M.P., India.
- Sanjeev Sharma is with Rajiv Gandhi Prodyogiki Vishwavidyalaya, Bhopal, M.P., India.

*Lack of fixed infrastructure:*
The absence of a fixed or central infrastructure is a key feature of MANETs. This eliminates the possibility to establish a centralized authority to control the network characteristics. Due to this absence of authority, traditional techniques of network management and security are scarcely applicable to MANETs.

*Resource constraints:*
MANETs are a set of mobile devices which are of low or limited power capacity, computational capacity, memory, bandwidth etc. by default. So in order to achieve a secure and reliable communication between nodes, these resource constraints make the task more enduring.

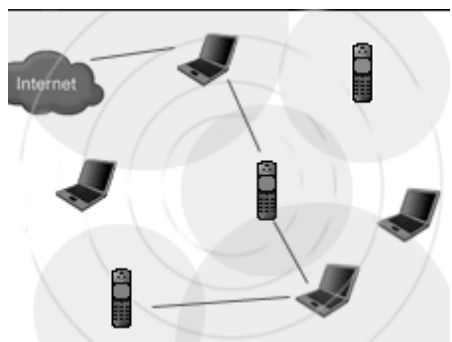

Fig. 1.    A typical MANET

Albeit the security requirements (availability, confidentiality, integrity, authentication, non-repudiation)[4] remain the same whether be it the fixed networks or MANETs, the MANETs are more susceptible to security attacks than fixed networks due their inherent characteristics.[5] Securitizing the routing process is a particular challenge due to open exposure of wireless channels and nodes to attackers, lack of central agency/infrastructure, dynamic topology etc.[6]. The wireless channels are accessible to all, whether meaningful network users or attackers with malicious intent. The lack of central agency inhibits the classical server based solutions to provide security. The dynamic topology entails that at any time any node whether legitimate or malicious can become a member of the network and disrupt the cooperative communication environment by purposely disobeying the routing protocol rules.

The rest of the paper is organized as follows: Section 2 presents routing protocols, Section 3 presents the presently known routing attacks, and Section 4 presents the various proposed countermeasures to these. Finally Section 5 summarizes the survey.

## 2 ROUTING PROTOCOLS IN MANETs

The nodes in MANETs perform the routing functions in addition to the inherent function of being the hosts. The limitation on wireless transmission range requires the routing in multiple hops. So the nodes depend on one another for transmission of packets from source nodes to destination nodes via the routing nodes. The nature of the networks places two fundamental requirements on the routing protocols. First, it has to be distributed. Secondly, since the topology changes are frequent, it should compute multiple, loop-free routes while keeping the communication overheads to a minimum. Based on route discovery time, MANET routing protocols fall into three general categories:
a) Proactive routing protocols
b) Reactive routing protocols
c) Hybrid routing protocols

### 2.1 PROACTIVE ROUTING PROTOCOLS

Proactive MANET protocols are table-driven and will actively determine the layout of the network. The complete picture of the network is maintained at every node, so route selection time is minimal. But the mobility of nodes if high then routing information in the routing table invalidates very quickly, resulting in many short lived routes. This also causes a large amount of traffic overhead generated when evaluating these unnecessary routes. For large size networks and the networks whose member nodes make sparse transmissions, most of the routing information is deemed redundant. Energy conservation being very important in MANETs, the excessive expenditure of energy is not desired.

Thus, proactive MANET protocols work best in networks that have low node mobility or where the nodes transmit data frequently. Examples of proactive MANET protocols include Optimized Link State Routing (OLSR)[7], Topology Broadcast based on Reverse Path Forwarding (TBRPF)[8], Fish-eye State Routing (FSR)[9], Destination-Sequenced Distance Vector (DSDV)[10], Landmark Routing Protocol (LANMAR)[11], Clusterhead Gateway Switch Routing Protocol (CGSR)[12].

### 2.2 REACTIVE ROUTING PROTOCOLS

Reactive MANET protocols only find a route to the destination node when there is a need to send data. The source node will start by transmitting route requests throughout the network. The sender will then wait for the destination node or an intermediate node (that has a route to the destination) to respond with a list of intermediate nodes between the source and destination. This is known as the global flood search, which in turn brings about a significant delay before the packet can be transmitted. It also requires the transmission of a significant amount of control traffic. Thus, reactive MANET protocols are most suited for networks with high node mobility or where the nodes transmit data infrequently. Examples of reactive MANET protocols include Ad Hoc On-Demand Distance Vector (AODV) [13], Dynamic Source Routing (DSR) [14], Temporally Ordered Routing Algorithm (TORA) [15], Dynamic MANET On Demand (DYMO) [16].

### 2.3 HYBRID ROUTING PROTOCOLS

Since proactive and reactive routing protocols each work best in oppositely different scenarios, there is good reason to develop hybrid routing protocols, which use a mix of both proactive and reactive routing protocols. These hybrid protocols can be used to find a balance between the proactive and reactive protocols.

The basic idea behind hybrid routing protocols is to use proactive routing mechanisms in some areas of the network at certain times and reactive routing for the rest of the network. The proactive operations are restricted to a small domain in order to reduce the control overheads and delays. The reactive routing protocols are used for locating nodes outside this domain, as this is more bandwidth-efficient in a constantly changing network. Examples of hybrid routing protocols include Core Extraction Distributed Ad Hoc Routing Protocol (CEDAR) [17], Zone Routing Protocol (ZRP) [18], and Zone Based Hierarchical Link State Routing Protocol (ZHLS) [19].

## 3 ROUTING ATTACKS IN MANETs

All of the routing protocols in MANETs depend on active cooperation of nodes to provide routing between the nodes and to establish and operate the network. The basic assumption in such a setup is that all nodes are well behaving and trustworthy. Albeit in an event where one or more of the nodes turn malicious, security attacks can be launched which may disrupt routing operations or create a DOS (Denial of Service)[20] condition in the network.

Due to dynamic, distributed infrastructure-less nature of MANETs, and lack of centralized authority, the ad hoc networks are vulnerable to various kinds of attacks. The challenges to be faced by MANETs are over and above to those to be faced by the traditional wireless networks. The accessibility of the wireless channel to both the genuine user and attacker make the MANET susceptible to both passive eavesdroppers as well as active malicious attackers.

The limited power backup and limited computational capability of the individual nodes hinders the implementation of complex security algorithms and key exchange mechanisms. There is always a possibility of a genuine trusted node to be compromised by the attackers and subsequently used to launch attacks on the network. Node mobility makes the network topology dynamic forcing frequent networking reconfiguration which creates more chances for attacks.

The attacks on MANETs can be categorized as active or passive. In passive attacks the attacker does not send any message, but just listens to the channel. Passive attacks are non disruptive but are information seeking, which may be critical in the operation of a protocol. Active attacks may either be directed to disrupt the normal operation of a specific node or target the operation of the whole network.

A passive attacker listens to the channel and packets containing secret information (e.g., IP addresses, location of nodes, etc.) may be stolen, which violates confidentiality paradigm. In a wireless environment it is normally impossible to detect this attack, as it does not produce any new traffic in the network.

The action of an active attacker includes; injecting packets to invalid destinations into the network, deleting packets, modifying the contents of packets, and impersonating other nodes which violates availability, integrity, authentication, and non-repudiation paradigm. Contrary to the passive attacks, active attacks can be detected and eventually avoided by the legitimate nodes that participate in an ad hoc network [21].

In [22], the authors have surveyed attacks on MANETs and their countermeasures on protocol layer wise criteria. In [23], B.Kannhavong et al. have surveyed newer attacks like flooding, black hole, link withholding, link spoofing, replay, wormhole, colluding misrelay and their countermeasures. In [24], [25] the authors have presented an overview of secure routing protocols (Authenticated routing for ad hoc networks (ARAN)[26], Ariadne[27], Secure AODV (SAODV)[28], Secure Efficient Ad hoc Distance vector routing protocol (SEAD)[29], Secure Routing Protocol (SRP)[30], Secure Link-State Protocol (SLSP) [31]) in MANETs. In this article, we will survey the current state of art of routing attacks and their security measures.

The first approach to develop security solutions is the understanding of potential threats. Supported by this threat analysis and capabilities of potential attackers, the well known routing attacks in MANETs are discussed.

*Flooding Attack*:[32]
*Routing Table Overflow:*
The attacker node floods the network with bogus route creation packets to fake (non-existing) nodes or simply sends excessive route advertisements to the network. The purpose is to overwhelm the routing-protocol implementations, by creating enough routes to prevent new routes from being created or to overwhelm the protocol implementation. Proactive routing protocols, as they create and maintain routes to all possible destinations are more vulnerable to this attack.
*Sleep Depravation:*
In sleep deprivation attack, the resources of the specific node/nodes of the network are consumed by constantly keeping them engaged in routing decisions. The attacker node continually requests for either existing or non-existing destinations, forcing the neighboring nodes to process and forward these packets and therefore consume batteries and network bandwidth obstructing the normal operation of the network.
*Impersonation Attack:*
The attacker nodes impersonates a legitimate node and joins the network undetectable, sends false routing information, masked as some other trusted node.
*Black Hole Attack:*
In this attack, the attacker node injects false route replies to the route requests claiming to have the shortest path to the destination node whose packets it wants to intercept. Once the fictitious route has been established the active route is routed through the attacker node. The attacker node is then in a position to misuse or discard any or all of the network traffic being routed through it.
*Node Isolation Attack* [37]:
The authors in this work have introduced an attack against the OLSR protocol. As implied by the name, the goal of this attack is to isolate a given node from communicating with other nodes in the network. The idea of this attack is that attacker(s) prevent link information of a specific node or a group of nodes from being spread to the whole network. Thus, other nodes who could not receive link information of these target nodes will not be able to build a route to these target nodes and hence will not be able to send data to these nodes.
*Routing Table Poisoning Attack:*
Different routing protocols maintain tables which hold information regarding routes of the network. In poisoning attacks, the attacker node generates and sends fictitious traffic, or mutates legitimate messages from other nodes, in order to create false entries in the tables of the participating nodes. Another possibility is to inject a RREQ packet with a high sequence number. This causes all other legitimate RREQ packets with lower sequence numbers to be deleted [33]. Routing table poisoning attacks can result in selection of non-optimal routes, creation of routing loops, bottlenecks and even partitioning certain parts of the network.
*Wormhole Attack:*

The wormhole attack involves the cooperation between two attacking nodes [34]. One attacker captures routing traffic at one point of the network and tunnels it to another point in the network that shares a private high speed communication link between the attackers, and then selectively injects tunnel traffic back into the network. The two colluding attacker can potentially distort the topology and establish routes under the control over the wormhole link.

*Location Disclosure Attack:*

In this attack, the privacy requirements of an ad hoc network are compromised. Through the use of traffic analysis techniques or with simpler probing and monitoring approaches an attacker is able to discover the location of a node, and the structure of the network.

*Rushing Attacks* [35]:

The attacker (initiator) node initiates a Route Discovery for the target node. If the ROUTE REQUESTs for this Discovery forwarded by the attacker are the first to reach each neighbor of the target, then any route discovered by this Route Discovery will include a hop through the attacker. That is, when a neighbor of the target receives the rushed REQUEST from the attacker, it forwards that REQUEST, and will not forward any further REQUESTs from this Route Discovery. When non-attacking REQUESTs arrive later at these nodes, they will discard those legitimate REQUESTs. As a result, the initiator will be unable to discover any usable routes (i.e., routes that do not include the attacker) containing at least two hops (three nodes).

*Blackmail:*

The attack incurs due to lack of authenticity and it grants provision for any node to corrupt other node's legitimate information. Nodes usually keep information of perceived malicious nodes in a blacklist. This attack is relevant against routing protocols that use mechanisms for the identification of malicious nodes and propagate messages that try to blacklist the offender. An attacker may fabricate such reporting messages and tell other nodes in the network to add that node to their blacklists and isolate legitimate nodes from the network [36].

*Snare Attack* [38]:

Lin et al. have proposed the snare attack, which relates to military specific applications. In a battlefield, a node could be physically compromised (say when the corresponding soldier is caught by the enemy). Afterwards, the compromised node could be used to lure a Very Important Node, (say the commander), into communicating with it. Since the adversary can easily intercept any transmission in the network through the compromised node, the adversary can identify the physical location of the VIN by tracing and analyzing some routes. After locating the VINs, the adversary will be able to launch a Decapitation Strike on those VINs as a short cut to win the battle.

*The Invisible Node Attack* [39]:

Andel et al. have defined the invisible node attack and proved it to be different from the existing attacks (man in the middle, masquerading, and wormhole) and established its uniqueness. They have defined it as In any protocol that depends on identification for any functionality, any node that effectively participates in that protocol without revealing its identity is an invisible node and the action and protocol impact is termed an INA. Discussing the effects of INA on different routing protocols, they have shown it to be a unsolvable attack so far.

## 4 SECURITY MEASURES AGAINST ROUTING ATTACKS IN MANETs

In this section, we will discuss the countermeasures against the routing attacks and secured routing protocols in MANETS.

*Solutions to the Flooding Attack:*

In [40], Yi et al. have proposed a simple mechanism to prevent the flooding attack in the AODV protocol. Here each node is to monitor its neighbors' RREQ. If the RREQ rate of any neighbor exceeds the predefined threshold, the node records the ID of this neighbor in a blacklist. All future RREQs from the blacklisted nodes are then dropped. But this approach has limitations that a flooding threshold has to be set below which the attack cannot be detected. Also if a genuine nodes ID is impersonated by a malicious node and a large number of RREQs, are broadcast, other nodes might put the ID of this legitimate node on the blacklist.

In [41], Desilva et al. have proposed an adaptive technique to mitigate the effect of a flooding attack in the AODV protocol. It uses a statistical analysis to detect malicious RREQ floods and avoid the forwarding of such packets. The approach to attack detection is similar to that in [40.] with the difference that instead of a fixed threshold, this approach determines the threshold based on a statistical analysis of RREQs. The key advantage of this approach is that it can reduce the impact of the attack for varying flooding rates.

In [42], Guo et al. have proposed a flow based detection mechanism against the flooding attacks in MANETs using the non-parameter CUSUM algorithm [43]. For the attacks when the source and destination node addresses are generated randomly for flooding (address spoofing), the authors have defined a detection feature as the percentage of new RREQ flows from the total RREQ flows, over a small time interval. This percentage over a period of time should remain stably, at a low level for normal network situation. For nonaddress spoofing attacks, where the flooding RREQ have same source and destination node addresses, the detection feature is defined as the percentage of RREQ with a fixed set of source and destination node addresses to the total RREQ flows over a small time interval. This percentage over a period of time should remain stably, at a low level for normal network situation. These percentage variables being random, CUSUM algorithm has been used to detect the threshold level for attack condition. The authors have used the DSR protocol for the case study.

In [44] V.Balakrishnan et al. have proposed an obligation based model Fellowship to mitigate the flooding and packet drop attacks in MANETs. They have defined Rate Limitation, Enforcement and Restoration as the model parameters of Fellowship. Trust or security protocols can be used over Fellowship to further enhance the efficiency and improve the security in MANETs.

*Solutions to the Blackhole Attack:*
In [45] Tamilsevan et al. have proposed that the requesting node without sending the DATA packets to the reply node at once waits for other replies with next hop details from the other neighboring nodes. After receiving the first request a timer is set in the 'TimerExpiredTable', for collecting the further requests from different nodes. The 'sequence number', and the time at which the packet arrives is stored in a 'Collect Route Reply Table' (CRRT). Now the 'timeout' value based on arriving time of the first route request are calculated. Now CRRT is checked for any repeated next hop node which if found, it is assumed the paths are correct or the chance of malicious paths is limited. If there is no repetition then any random route from CRRT is selected.

In [46] Lee et al. have proposed the route confirmation request (CREQ) and route confirmation reply (CREP) to avoid the blackhole attack. The intermediate nodein addition to sending RREPs to the source node also sends CREQs to its next-hop node towards the destination node. The next-hop node on receipt of a CREQ looks up its cache for a route to the destination. If a route is found, it sends the CREP to the source. On receipt of the CREP, the source node compares the path in RREP and the one in CREP. If both are identical the source node pronounces the route to be correct. However in this proposal a blackhole attack is not resolved if two consecutive nodes work in collusion, that is, when the next-hop node is a colluding attacker.

In [47], Shurman et al. have proposed the source node to wait until the arrival of a RREP packet from more than two nodes. On receiving multiple RREPs, the source node checks about a shared hop. If at least one hop is shared, the source node judges that the route is safe. The drawback here is the introduction of a time delay due to the wait till the arrival of multiple RREPs.

In [48], Kurosowa et al. have analyzed the blackhole attack and propounded that the destination sequence number must sufficiently be increased by the attacker node in order to convince the source node that the route provided is optimum. Based on differences between the destination sequence numbers of the received RREPs, the authors propose a statistical based anomaly detection approach to detect the blackhole attack. This approach has a merit that the attack can be detected at a low cost without introducing extra routing traffic without modification of the existing protocol, albeit false positives is a demerit.

*Solution to Node Isolation Attack:*
In [37] Kannhavong et al have shown that a malicious node can isolate a specific node and prevent it from receiving data packets from other nodes by withholding a TC message in OLSR protocol. A detection technique based on observation of both a TC message and a HELLO message generated by the MPR nodes is proposed. If a node does not hear a TC message from its MPR node regularly but hears only a HELLO message, a node judges that the MPR node is suspicious and can avoid the attack by selecting one or more extra MPR nodes.

In [49], Dillon et al. have proposed an IDS that detects TC link and message withholding in the OLSR protocol. Each node is set to observe whether a MPR node generates a TC message regularly or not. If a MPR node generates a TC message regularly, the node checks whether or not the TC message actually contains itself to detect the attack. The draw back of these approaches is that they cannot detect the attack if it that is launched by two colluding next hop nodes, where the first attacker pretends to advertise a TC message, but the second attacker drops this TC message.

*Solutions to the Worm Hole Attack:*
In [50], packet leashes are proposed to detect and defend against the wormhole attack. Hu et al. in their work have proposed temporal leashes and geographical leashes. For temporal leashes each node is to compute the packet expiration time ($t_e$) based on the speed of light c and is to include the expiration time ($t_e'$) in its packet to prevent the packet from traveling further than a specific distance, L. At the receiving node, the packet is checked for packet expiry by comparing its current time and the $t_e$ in the packet. The authors also proposed TIK, which is used to authenticate the expiration time that can otherwise be modified by the malicious node. The constraint here is that all nodes have to be tightly clock synchronized. For the geographical leashes, each node must know its own position and may have loosely synchronized clocks. In this approach, a sender of a packet includes its current position and the sending time. Therefore, a receiver can judge neighbor relations by computing distance between itself and the sender of the packet. The advantage of geographic leashes over temporal leashes is that the time synchronization is not critical.

In [51] Qian et al. have proposed a Statistical Analysis of Multipath (SAM), which is an approach to detect the wormhole attack by using multipath routing. The attack is detected by calculating the relative frequency of each link that appears in all of the obtained routes from one route discovery. The link that has the highest relative frequency is identified as the wormhole link.

In [52] Su et al. have proposed technique based on propagation speeds of requests and statistical profiling. For on demand route discovery schemes that use flooding, requests should be transmitted at a higher priority than all other packets. This implicitly increases the time to exchange information among malicious nodes. A distributed and adaptive statistical profiling technique to filter RREQs (each destination node filters RREQs that are targeted to it and have excessively large delays) or RREPs (each source node monitors the RREPs it receives and filters those that have excessively large delays) is suggested. Since different RREQs/RREPs take varying number of hops, the upper bound on the per hop time of RREQ/RREP packets is so calculated that most normal packets are retained and most falsified packets are filtered. The main advantages of this approach are that no network-wide synchronized clocks are required, no additional control packet overhead is imposed and only simple computations by the sources or destinations of connections is required.

In [53] Gorlatova et al. have proposed an approach that uses the anomaly in the MANET traffic behavior, particularly the behavioral anomalies in the protocol related packets for detection of worm holes. The HELLO message interval was set to 0.3 seconds, with a simple jitter function - randomly adding 0.03 seconds of delay overlaid upon it. The traffic is parsed, the HELLO messages arriving at a particular node are indexed, and the difference between arrival times of HELLO messages sent by its neighbors is calculated. The HELLO Message Timing Interval HMTI profile so obtained is used for detection of attacker nodes, as the frequency profile of HMTI is at a set frequency, a violation of OLSR protocol specifications. The interval between the packets is repeatedly much larger than it should be for a genuine node.

*Solutions to the Rushing Attack:*
In [35] Hu et al. have proposed a set of generic mechanisms that together defend against the rushing attack: Secure neighbor detection, Secure route delegation, and Randomized ROUTE REQUEST forwarding. Secure neighbor detection allows each neighbor to verify that the other is within a given maximum transmission range. Once a node A determines that node B is a neighbor it signs a Route Delegation message, allowing node B to forward the ROUTE REQUEST. When node B determines that node A is within the allowable range, it signs an Accept Delegation message. The Randomized selection of ROUTE REQUEST message to be forwarded, which replaces traditional duplicate suppression in on-demand route discovery, ensures that paths that forward REQUESTs with low latency are only slightly more likely to be selected than other paths.

*Solution to the Snare Attack:*
In [38] Lin et al. have defined the snare attack, and proposed ASRPAKE (An Anonymous Secure Routing Protocol with Authenticated Key Exchange for Wireless Ad Hoc Networks) and Decoy node deployment to mitigate this attack. The proposed anonymous secure routing protocol consists of five phases: the key pre-distribution phase, the neighborhood discovery phase, the route discovery phase, the route reverse phase, and the data forwarding phase. The anonymity of the VIN can further be enhanced using n no. of decoy nodes which allow the communication to be routed to the VIN only after verifying the authenticity of the source node. The main features of this approach are achievable end-to-end anonymity and security, and the integration of the authenticated key exchange operations into the routing algorithm.

*Trust Based Security Solutions:*
Another active area of research in Mobile Ad Hoc and Sensor Network security in general is the Trust Based Security Solutions. In [54] Sun et al. have identified the role of Trust in MANETs. When a network entity establishes trust in other network entities, it can predict the future behaviors of others and diagnose their security properties. Trust helps in Assistance in decision making to improve security and robustness, Adaptation to risk leading to flexible security solutions, Misbehavior detection and Quantitative assessment of system-level security properties.

Balakrishnan et al. in [44], [55],[56] have done extensive work on Trust based security solutions and have proposed Fellowship, TEAM (Trust Enhanced Security Architecture for Mobile Ad-hoc Networks) SMRITI (Secure MANET Routing with Trust Intrigue). In TEAM a trust model (SMRITI) is overlaid on other security models such as key management, secure routing and cooperation model (Fellowship) to enhance security. SMRITI assists the security models in making routing decisions, corresponding to the Trust evaluation of the involved nodes. The advantage of this approach is that no special/tamper proof hardware is required and there is no requirement of a central authority as well.

## 5 SUMMARY

MANETs is an emerging technological field and hence is an active area of research. Because of ease of deployment and defined infrastructure less feature these networks find applications in a variety of scenarios ranging from emergency operations and disaster relief to military service and task forces. Providing security in such scenarios is critical.

The primary limitation of the MANETs is the limited resource capability: bandwidth, power back up and computational capacity. Absence of infrastructure, vulnerability of channels and nodes, dynamically changing topology make the security of MANETs particularly difficult. Also no centralized authority is present to monitor the networking operations. Therefore, existing security schemes for wire networks cannot be applied directly to a MANETs, which makes them much more vulnerable to security attacks.

Of these attacks, the passive attacks do not disrupt the operation of a protocol, but is only information seeking in nature whereas active attacks disrupt the normal operation of the MANET as a whole by targeting specific node(s).

In this survey, we reviewed the current state of the art routing attacks and countermeasures MANETs. The advantages as well as the drawbacks of the countermeasures have been outlined.

It has been observed that although active research is being carried out in this area, the proposed solutions are not complete in terms of effective and efficient routing security. There are limitations on all solutions. They may be of high computational or communication overhead (in case of cryptography and key management based solutions) which is detrimental in case of resource constrained MANETS, or of the ability to cope with only single malicious node and ineffectiveness in case of multiple colluding attackers. Some solutions may require special hardware such as a GPS or a modification to the existing protocol. Furthermore, most of the proposed solutions can work only with one or two specific attacks and are still vulnerable to unexpected attacks.

A number of challenges like the Invisible Node Attack remain in the area of routing security of MANETs. Although researchers have designed efficient security routing, optimistic approaches like Fellowship-TEAM-SMRITI [44, 55, 56], CREQ-CREP approach [45] etc., which can provide a better tradeoff between security and performance, a lot more is yet to be done. Future research efforts should be focused not only on improving the effectiveness of the security schemes but also on minimizing the cost to make them suitable for a MANET environment.